\documentclass[page-classic,showpacs]{epl2}

\usepackage{graphicx}
\usepackage[centertags]{amsmath}

\newcommand{\cm}{\ensuremath{\,\mbox{cm}^{-1}}}
\newcommand{\K}{\ensuremath{\,\mbox{K}}}
\newcommand{\celsius}{\ensuremath{\,{}^\circ}\!C}

\begin{document}
\title{Magnetodielectric effect and optic soft mode behaviour in quantum paraelectric EuTiO$_{3}$ ceramics}
\shorttitle{Magnetodielectric effect in quantum paraelectric EuTiO$_{3}$}

\author{S.~Kamba\inst{1} \and D.~Nuzhnyy\inst{1} \and P. Van\v{e}k\inst{1} \and M. Savinov\inst{1} \and K. Kn\'{i}\v{z}ek\inst{1}
\and Z. Shen\inst{2} \and E. \v{S}antav\'{a}\inst{1} \and K. Maca\inst{3} \and
M.~Sadowski\inst{4} \and J.~Petzelt\inst{1}}
 \shortauthor{S.~Kamba \etal}

\institute{
  \inst{1} Institute of Physics ASCR, v.v.i. Na Slovance~2, 182 21 Prague~8, Czech
Republic\\
  \inst{2} Arrhenius Laboratory, Stockholm University, SE-10691 Stockholm\\
  \inst{3} Brno University of Technology, Technicka 2896/2, 616 69 Brno, Czech
Republic\\
  \inst{4} Grenoble High Magnetic Field Lab, CNRS, 25, avenue des Martyrs, Grenoble
Cedex 9, France
}


\abstract{ Infrared reflectivity and time-domain terahertz
transmission spectra of EuTiO$_{3}$ ceramics revealed a polar
optic phonon at 6 - 300\K\, whose softening is fully responsible
for the recently observed quantum paraelectric behaviour. Even if
our EuTiO$_{3}$ ceramics show lower permittivity than the single
crystal due to a reduced density and/or small amount of secondary
pyrochlore Eu$_{2}$Ti$_{2}$O$_{7}$ phase, we confirmed the
magnetic field dependence of the permittivity, also slightly
smaller than in single crystal. Attempt to reveal the soft phonon
dependence at 1.8\,K on the magnetic field up to 13\,T remained
below the accuracy of our infrared reflectivity experiment.}

\maketitle

 Multiferroics exhibiting simultaneously ferroelectric and ferro- or
antiferromagnetic order are known since the beginning of the 1960's, but the interest to
these materials underwent a revival after the pioneering work by Wang et al.\cite{wang03}
who in the BiFeO$_{3}$ thin films revealed spontaneous polarization and magnetization
almost by an order of magnitude higher compared to the bulk samples. Recently, much
attention have been paid to the multiferroic materials not only because of their rich and
fascinating fundamental physics (questions why only few materials with simultaneous
ferroelectric and magnetic order are known, how to explain the coupling of magnetic and
ferroelectric order etc.), but also because of the promising potential applications in
multiple-state memory elements.\cite{fiebig05,cheong07} In magnetoelectrics the
polarization can be controlled by the magnetic field as well as the magnetization by the
electric field and so magnetic field tuning of the dielectric permittivity $\varepsilon$'
(magnetodielectric effect) is expected. Recently, gigantic proper magnetodielectric
effect has been observed e. g. in TbMnO$_{3}$\cite{kimura03} and
EuTiO$_{3}$.\cite{katsufuji01} Large magnetodielectric effect was recently seen also in
BiFeO$_{3}$\cite{kamba07}, however in this case the change of $\varepsilon$' with
magnetic field is not due to coupling of the polarization and magnetization, but due to
combination of inhomogeneous magnetoresistance and Maxwell-Wagner effect.\cite{kamba07}

The perovskite EuTiO$_{3}$ is not a typical multiferroic. It exhibits an
antiferromagnetic (AFM) structure below $T_{N}$= 5.3\K,\cite{guire66} but the
ferroelectric order does not take place because the quantum fluctuations prevent the
freezing of polarization at low temperatures. Its permittivity increases on cooling
similarly to classical quantum paraelectrics\cite{mueler79,samara01} (incipient
ferroelectrics) SrTiO$_{3}$, KTaO$_{3}$, etc., but it saturates below $\sim$30\K\, and
sharply drops down at $T_{N}$.\cite{katsufuji01} Magnetic structure of EuTiO$_{3}$ was
investigated already forty years ago\cite{guire66} and determined in Ref. \cite{chien74}.
A neutron-diffraction study of a powder sample revealed the G type AFM
structure.\cite{guire66} In this magnetic arrangement, there are two interpenetrating
$fcc$ sublattices in which a given Eu$^{2+}$ has six nearest-neighbor Eu$^{2+}$ with
antiparallel spins and 12 next-nearest-neighbor Eu$^{2+}$ with parallel spins. At
1.3\,K\, the magnetic moment increases linearly with magnetic field up to 1\,T and above
1.4\,T the moment saturates at 156 emu/g (6.93 $\mu_{B}$).\cite{guire66}

The low-temperature $\varepsilon$' below $T_{N}$ exhibits a huge
dependence on the magnetic field $B$, giving evidence about a
large interaction of magnetic moments with the crystal
lattice.\cite{katsufuji01} $\varepsilon$' strongly increases even
for low $B$ and the drop down in $\varepsilon$' seen near $T_{N}$
at 0\,T disappears at fields above 1\,T so that only quantum
paraelectric behaviour is seen at high magnetic fields down to
2\K. Katsufuji and Takagi\cite{katsufuji01} suggested that the
temperature dependence of $\varepsilon$' above $T_{N}$ is due to a
soft optic phonon reducing its frequency on cooling and the
decrease in $\varepsilon$' below $T_{N}$ occurs due to the strong
coupling of localized spins in the $4f$ levels of Eu$^{2+}$ with
the soft $F_{1u}$ phonon, which causes its hardening below
$T_{N}$. The transverse-field Ising model and Heisenberg model
were successfully applied for explaining the dielectric and
magnetic properties of EuTiO$_{3}$ and Eu$_{1-x}$Ba$_{x}$TiO$_{3}$
with and without external magnetic and electric
fields.\cite{jiang03,gong04,wu04a,wu04b,wu05} Jiang and
Wu\cite{jiang03} calculated the soft mode frequency and obtained
8\cm\, at 20\K. Fennie and Rabe\cite{fennie06} calculated the soft
mode frequency in the AFM phase from the first principles and
obtained 77\cm. They also predicted a remarkable softening of the
soft mode and transition into the ferroelectric phase in biaxially
strained samples. The strain can be realized in thin films
deposited on substrates with slightly mismatched lattice
parameters.\cite{fennie06} In such a phase the giant
magnetodielectric effect is expected.

The polar soft mode is infrared (IR) active so that the IR spectroscopy is the best tool
for the verification of the predicted phonon softening on cooling. In this letter we will
show that the soft mode really exhibits the expected softening (from 112\cm\, at 300 K to
82\cm\, at 6K), in perfect agreement with the predictions from the first-principles
calculations.\cite{fennie06}

EuTiO$_{3}$ ceramics was synthesized from the Eu$_{2}$O$_{3}$ and Ti$_{2}$O$_{3}$ powders
using mechanochemical activation in the planetary micromill Fritsch Pulverisette 7.
Powder XRD showed strong amorphization - no diffraction line of Eu$_{2}$O$_{3}$ were
found but significantly widened peaks of the Ti$_{2}$O$_{3}$ and of some unidentified
phases. The powder was pressed into pellet in a uniaxial press at 650\,MPa (or in an
isostatic press at 300\,MPa) at room temperature and then annealed at 1500\celsius\, in
Ar + 10\%\,H$_{2}$ atmosphere. The reducing atmosphere is necessary to prevent the
formation of pyrochlore Eu$^{3+}_{2}$Ti$^{4+}_{2}$O$_{7}$. The powder XRD showed sharp
diffraction lines corresponding to single-phase cubic perovskite EuTiO$_{3}$. No other
phase was detected, but the porosity of resulting ceramics was relatively high - 20-30\%.
Therefore we used an additional sample processing: The EuTiO$_{3}$ pellet was ground and
milled again to a fine nanopowder which was sintered by spark plasma sintering
(temperature 1150-1200\celsius, pressure 75-100\,MPa, time 3-5\,min). The resulting
ceramic samples were more than 91\% dense and contained 5\% or in one case even 15\% of
the secondary pyrochlore phase.

The magnetodielectric effect was studied by measuring the changes
in permittivity with the magnetic fields up to 14\,T (PPMS,
Quantum design) at temperatures 2-300\K. The measurements were
performed at frequency 1\,kHz with ultra-precision capacitance
bridge Andeen-Hagerling 2500A. Details of the dielectric, THz and
IR experiments performed without magnetic field are described
elsewhere.\cite{kamba07} The IR reflectivity spectra were taken as
well at magnetic fields up to 13\,T at 1.8\K\, using a Fourier
transform IR spectrometer Bruker IFS 113v. For reduction of the
high noise we measured 2000 scans with resolution of 6\cm, while
the IR spectra in non-magnetic cryostat were taken with resolution
of 2\cm\, at only 128 scans.

\begin{figure}
  \begin{center}
    \includegraphics[width=80mm]{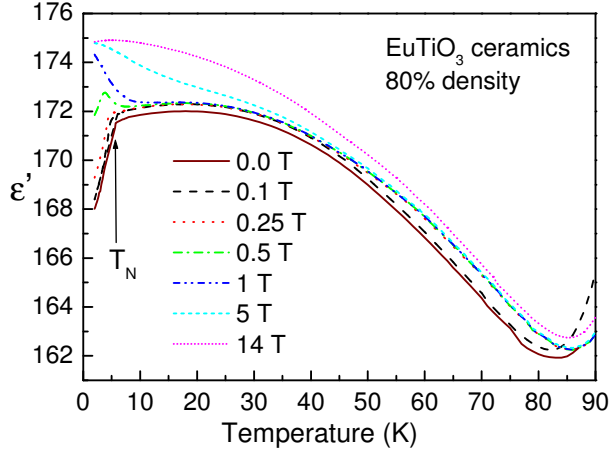}
  \end{center}
    \caption{(color online) Temperature dependence of the permittivity in EuTiO$_{3}$
    ceramics
    measured at 1\,kHz and at various magnetic fields. Critical temperature to AFM phase is marked.}
    \label{Fig1}
\end{figure}

Temperature dependence of $\varepsilon$' below 90\K\, measured at various magnetic fields
is plotted in Fig.~\ref{Fig1}. At higher temperature the sample becomes slightly
conducting,\cite{katsufuji99} therefore $\varepsilon$' increases to huge values (not
shown in Fig.~\ref{Fig1}) due to Maxwell-Wagner polarization mechanism. Below 85\,K, one
can see an increase in intrinsic permittivity on cooling and its saturation below $\sim$
25\K\, due to quantum fluctuations. The shape of the $\varepsilon$'($T$) curve down to
6\K\, is the same as in classical quantum paraelectric SrTiO$_{3}$,\cite{mueler79} but
with much lower value. Sharp drop appears in $\varepsilon$'($T$) below AFM phase
transition at $T_{N}$=5.3\K, although the sample remains paraelectric. At high magnetic
fields above 1\,T the AFM phase transforms to the ferromagnetic one, drop down in
$\varepsilon$'($T$) disappears and a pure incipient ferroelectric behaviour, i.e.
continuous increase of $\varepsilon$'(T) on cooling is seen. We note that the same
temperature and magnetic field dependences of $\varepsilon$' were observed independently
on orientation of magnetic $\textbf{B}$ and measuring electric $\textbf{E}$ fields (i.e.
both geometries $\textbf{E}\| \textbf{B}$ and $\textbf{E} \bot \textbf{B}$ gave the same
$\varepsilon$'($T,\textbf{B}$) results as in Fig.~\ref{Fig1}). Similar
$\varepsilon$'($T,\textbf{B}$) was observed by Katsufuji and Takagi on EuTiO$_{3}$ single
crystal, but the value of permittivity was more than twice higher compared with our
values obtained on the ceramics of 80\% density. This discrepancy could be due to the
porosity which may cause a 50\% decrease in the permittivity.\cite{rychetsky02}

\begin{figure}
  \begin{center}
    \includegraphics[width=83mm]{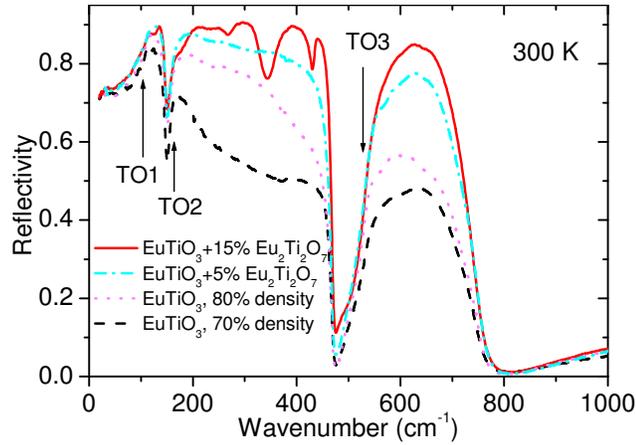}
  \end{center}
    \caption{(color online) Room-temperature IR reflectivity spectra of EuTiO$_{3}$ ceramics of 70 and 80 \% density as well as
    EuTiO$_{3}$ + 15\% Eu$_{2}$Ti$_{2}$O$_{7}$ (93.6\% density) and  EuTiO$_{3}$ + 5\% Eu$_{2}$Ti$_{2}$O$_{7}$ (91\%
    density). Transverse phonon frequencies of three perovskite modes are marked.}
    \label{Fig2}
\end{figure}

$\varepsilon$' decreases with the AFM ordering of Eu spins whereas it increases with
their ferromagnetic arrangement under magnetic field. Therefore Katsufuji and Takagi
suggested that the $\varepsilon$'($\textbf{B}$) is dominated by the pair correlation of
the Eu neighbouring spins $\langle \textbf{S}_{i}\cdot \textbf{S}_{j}\rangle$ and
successfully fitted the experimental $\varepsilon$'($T$) data by the formula
\begin{equation}\label{spin}
\varepsilon'(T,\textbf{B})=\varepsilon'_{0}(T)(1+\alpha\langle \textbf{S}_{i}\cdot
\textbf{S}_{j}\rangle),
\end{equation}
where $\varepsilon'_{0}$($T$) is the dielectric constant in the
absence of a spin correlation and $\alpha$ is the coupling
constant between correlated spins and permittivity. The same
authors also suggested that the value of static permittivity is
only due to contributions of polar optic phonons and the
temperature dependence of $\varepsilon'_{0}$ is caused by
softening of one of the phonons, similarly as in SrTiO$_{3}$. Our
radio-frequency dielectric measurements without magnetic field
really did not reveal any frequency dispersion in $\varepsilon'$
and only negligible dielectric losses between 100\,Hz and 1\,MHz.

\begin{figure}
  \begin{center}
    \includegraphics[width=85mm]{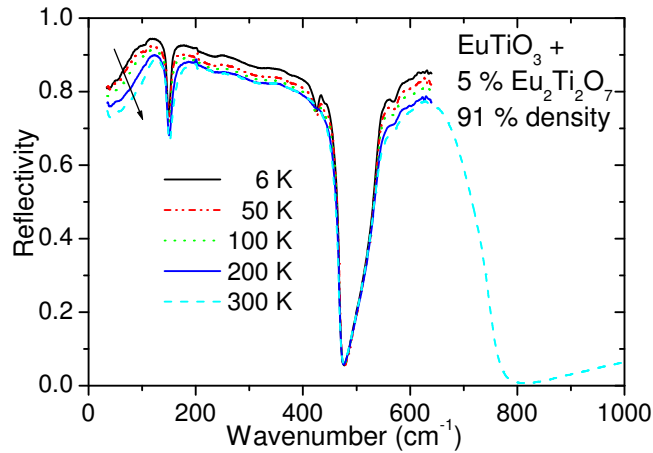}
  \end{center}
    \caption{(color online) IR reflectivity spectra of the EuTiO$_{3}$ ceramics at various temperatures.
    Increase in temperature is marked by the arrow. Softening of the low-frequency phonon on cooling is clearly
    seen. Note that 1\cm\, = 30\,GHz.}
    \label{Fig3}
\end{figure}

\begin{figure}
  \begin{center}
    \includegraphics[width=70mm]{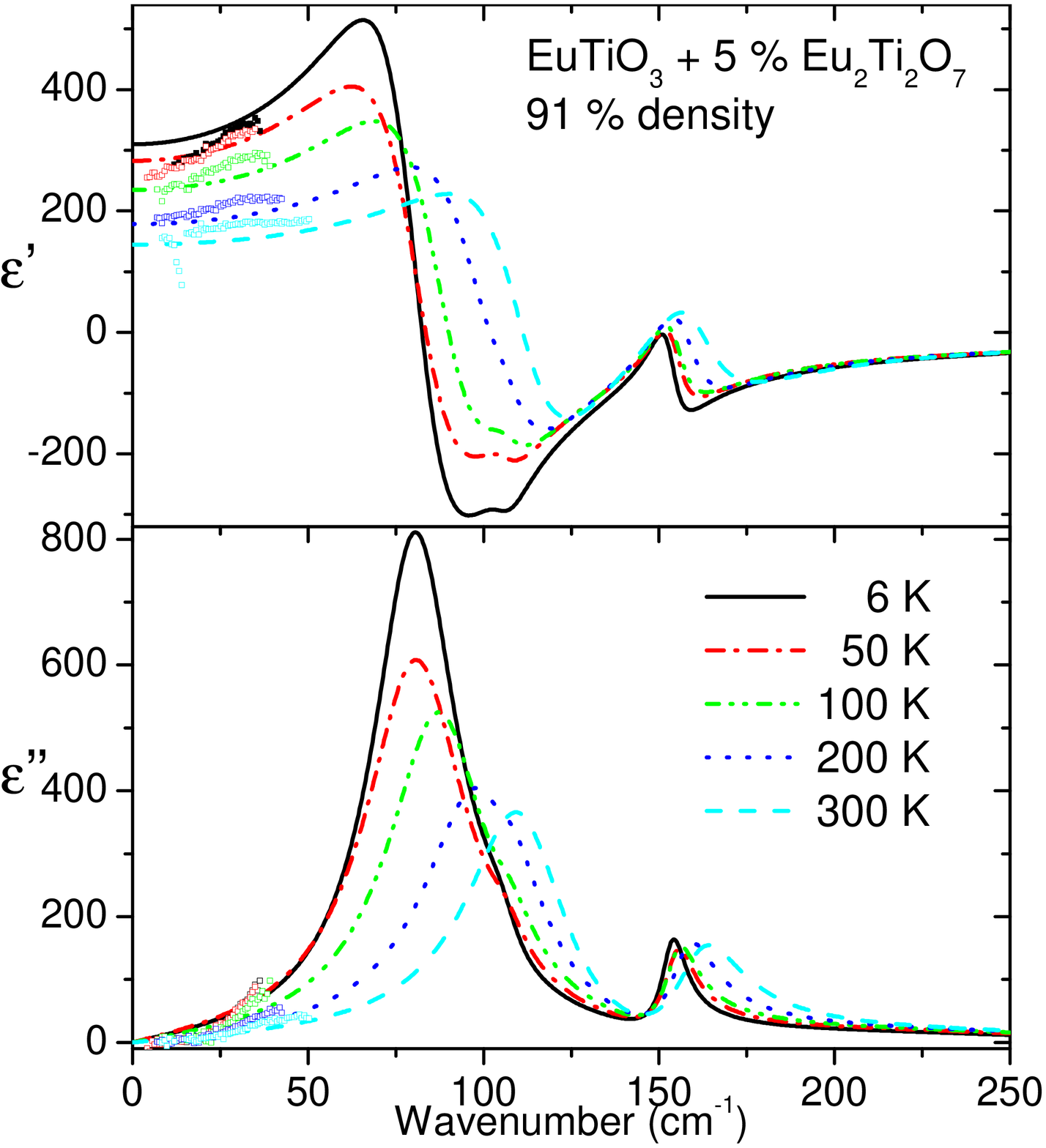}
  \end{center}
    \caption{(color online) Complex dielectric spectra of EuTiO$_{3}$ ceramics obtained from the fits of the IR reflectivity at
    selected temperatures. Spectra above 250\cm\, are not plotted since they
    show negligible temperature dependence. Dots correspond to the data evaluated from the THz spectra.}
    \label{Fig4}
\end{figure}

For confirmation of the anomalous polar phonon we performed THz
transmission and IR reflectivity measurements between 6 and 300\K.
IR spectra show strong dependence on the density of the ceramics
as well as on the amount of pyrochlore Eu$_{2}$Ti$_{2}$O$_{7}$
secondary phase (see Fig.~\ref{Fig2}). Low-dense pure perovskite
ceramics shows a large diffuse scattering, which rises with
increasing frequency and deteriorates the IR reflectivity. 93.6\%
dense EuTiO$_{3}$ ceramics with 15\% of Eu$_{2}$Ti$_{2}$O$_{7}$
secondary phase exhibits no diffuse scattering but additional
reflection bands between 100 and 450\cm, corresponding to polar
phonons of the pyrochlore phase, while in the perovskite cubic
phase only 3 IR active phonons are allowed. As an optimal
compromise we decided to present here the low-temperature spectra
of the sample with 5\% of pyrochlore phase (91\% density) which
exhibits only weak diffuse scattering and the bands from
pyrochlore phase are also very weak (see Figs.~\ref{Fig2}
and~\ref{Fig3} ).

Fig.~\ref{Fig3} shows corresponding IR reflectivity spectra at
selected temperatures. One can see an increase of reflectivity on
cooling due to the reduced phonon damping at low temperatures, as
well as the shift of the first reflection band (i.e. the soft
mode) to lower frequencies on reducing temperature. Complex
permittivity
$\varepsilon$$^{*}$($\omega)=\varepsilon'(\omega)-\textrm{i}\varepsilon''(\omega)$
is related to the reflectivity R($\omega$) by
\begin{equation}\label{refl}
R(\omega)=\left|\frac{\sqrt{\varepsilon^{*}(\omega)}-1}{\sqrt{\varepsilon^{*}(\omega)}+1}\right|^2.
\end{equation}
For the fit of the IR and THz spectra we used a generalized-oscillator model with the
factorized form of the complex permittivity:\cite{gervais83}
\begin{equation}\label{eps4p}
\varepsilon^{*}(\omega)=\varepsilon_{\infty}\prod_{j}\frac{\omega^{2}_{LOj}-\omega^{2}+\textrm{i}\omega\gamma_{LOj}}
{\omega^{2}_{TOj}-\omega^{2}+\textrm{i}\omega\gamma_{TOj}}
\end{equation}
where $\omega_{TOj}$ and $\omega_{LOj}$ denotes the transverse and longitudinal frequency
of the j-th polar phonon, respectively, and $\gamma$$_{TOj}$ and $\gamma$$_{LOj}$ denotes
their corresponding damping constants. The high-frequency permittivity
$\varepsilon_{\infty}$ resulting from the electron absorption processes was obtained from
the room-temperature frequency-independent reflectivity tail above the phonon frequencies
and was assumed temperature independent.

Real and imaginary parts of $\varepsilon^*$($\omega$) obtained
from the fits to IR reflectivity are shown together with the
experimental THz spectra in Fig.~\ref{Fig4}. The maxima in
$\varepsilon''$($\omega$) correspond roughly to the phonon
eigen-frequencies. One can clearly see the increase in static
permittivity on cooling (zero frequency values in
Fig.~\ref{Fig4}a) due to the phonon softening.

It is worth to note that for the fits we used the factorized form
of $\varepsilon^{*}$ (Eq.~(\ref{eps4p})) instead of more
frequently used classical damped harmonic oscillators
model\cite{gervais83}

\begin{equation}
\label{eps3p}
 \varepsilon^*(\omega)
 = \varepsilon_{\infty} + \sum_{j=1}^{n}
\frac{\Delta\varepsilon_{j}\omega_{TOj}^{2}} {\omega_{TOj}^{2} -
\omega^2+\textrm{i}\omega\gamma_{TOj}} \, ,
\end{equation}

where dielectric strength $\Delta\varepsilon_{j}$ means the contribution of the j-th mode
to the static permittivity and the rest of the parameters in Eq.~(\ref{eps3p}) has the
same meaning as in Eq.~\ref{eps4p}. The Eq.~\ref{eps4p} is more suitable than
Eq.~(\ref{eps3p}) for the fits of IR reflectivity spectra of EuTiO$_{3}$, because it
exhibits large TO-LO splitting, which cannot be well fitted with Eq.~(\ref{eps3p}).
Dielectric strength $\Delta\varepsilon_{j}$ can be obtained from the
formula\cite{gervais83}

\begin{equation}
 \Delta\varepsilon_{j} = \varepsilon_{\infty}\omega^{-2}_{TOj}\frac{\prod_{k}\omega^{2}_{LOk}-\omega^{2}_{TOj}}{\prod_{k\neq
 j}\omega^{2}_{TOk}-\omega^{2}_{TOj}}.
 \label{eq:sila}
\end{equation}

\begin{figure}
  \begin{center}
    \includegraphics[width=85mm]{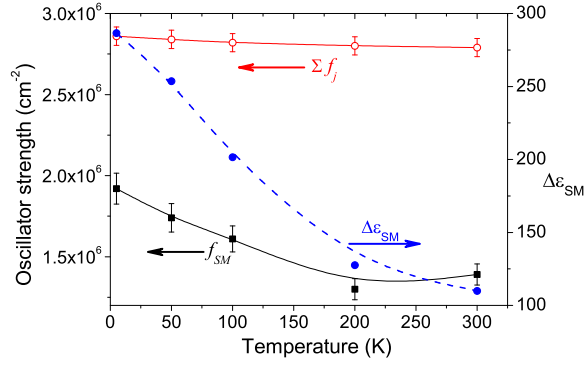}
  \end{center}
    \caption{(color online) Temperature dependence of oscillator strength $f_{SM}=\Delta\varepsilon_{SM}\omega_{SM}^{2}$ of the lowest
    frequency soft phonon, the sum of all oscillator strengths as well as dielectric strength $\Delta\varepsilon_{SM}$
    of the soft mode.
    Note different right scale for $\Delta\varepsilon_{SM}(T)$. }
    \label{Fig5}
\end{figure}

It follows from very general summation rules\cite{smith85} that
the sum $f$ of all oscillator strengths
$f_{j}=\Delta\varepsilon_{j}\omega_{TOj}^{2}$ is temperature
independent (i.e. $f(T)=\sum_{j=1}^n{f_j}(T) = const.$). In the
case of uncoupled phonons even each oscillator strength $f_{j}$
should remain temperature independent (i.e.
$\Delta\varepsilon_{j}(T).\omega_{TOj}^2(T) = const.$). It means
that each decrease of phonon frequency $\omega_{TOj}$ should be
accompanied by increase of dielectric strength
$\Delta\varepsilon_{j}$. Fig.~\ref{Fig5} shows that the dielectric
strength $\Delta\varepsilon_{SM}$ of the lowest frequency mode
(i.e. the soft mode) remarkably increases on cooling. However, the
oscillator strength $f_{SM}$ of the soft mode increases on cooling
as well, which gives evidence about the coupling of the soft mode
with other higher-frequency modes. It means that the soft modes
receives some part of its dielectric strength from
higher-frequency modes, predominantly from second perovskite (TO2)
phonon (see in Fig.~\ref{Fig4} the shift of second peak in
$\varepsilon''$ with temperature). Simultaneously, the
Fig.~\ref{Fig5} shows that the sum rule of oscillator strengths is
valid, because $\sum_{j=1}^n{f_j}(T) = const.$ within accuracy of
our measurements.

EuTiO$_{3}$ crystallizes in the cubic $Pm\bar{3}m$ structure, which allows 3$F_{1u}$ IR
active phonons and no Raman active mode. Our Raman spectra really revealed no first order
peak. In IR spectra we see three distinct reflection bands corresponding to phonon
frequencies at 82, 153 and 539\cm\, (at 6\K), however the fits were performed on the
whole with 12 (mostly weak) modes, which account for the small rippled shape of
reflectivity (mostly between 100 and 450\cm). The additional modes stem apparently from
the secondary pyrochlore Eu$_{2}$Ti$_{2}$O$_{7}$ phase since they are better resolved in
the sample with 15\% pyrochlore phase (Fig.~\ref{Fig2}). We studied also temperature
dependence of IR reflectivity of the 80\%-dense single phase perovskite sample used for
the magnetodielectric experiment in Fig.~\ref{Fig1}. The spectra yield the $\textit{same
temperature dependence of the soft mode}$, however, we do not present the spectra because
of the deteriorated reflectance above 200\cm.

\begin{figure}
  \begin{center}
    \includegraphics[width=85mm]{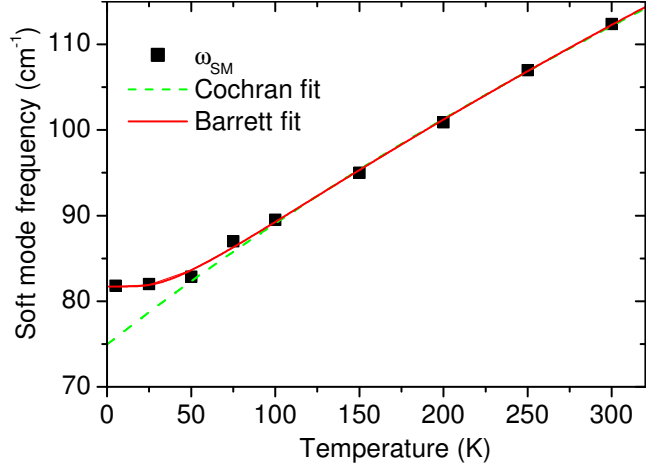}
  \end{center}
    \caption{(color online) Temperature dependence of the soft mode frequency fitted with Cochran and Barrett formula.}
    \label{Fig6}
\end{figure}

\begin{figure}
  \begin{center}
    \includegraphics[width=85mm]{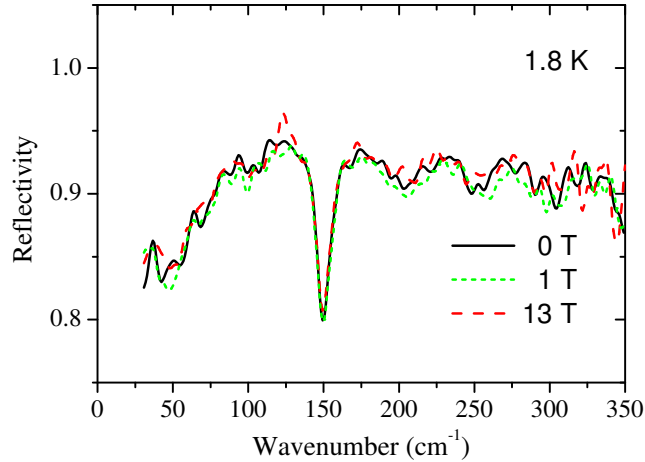}
  \end{center}
    \caption{(color online) FIR reflectivity spectra of EuTiO$_{3}$ taken at 1.8\K\, and at various magnetic fields.
    No spectral change with magnetic field is seen.}
    \label{Fig7}
\end{figure}

In Fig.~\ref{Fig6} the temperature dependence of the soft mode
frequency $\omega_{SM}$ is plotted. The soft mode softens from
112\cm\, (300\K) down to 82\cm\, (6\K) and below 50\K\, is seen
the saturation of $\omega_{SM}(T)$. The low-temperature value of
$\omega_{SM}$ reasonably corresponds to the soft mode frequency
obtained from the first principle calculations by Fennie and
Rabe,\cite{fennie06}, but is one order of magnitude larger than
the value estimated by Jiang and Wu\cite{jiang03}. In classical
paraelectrics above the ferroelectric phase transition $T_{C}$,
the $\omega_{SM}$ should obey the Cochran law
$\omega_{SM}=\sqrt{A(T-T_{C})}$. Our fit in Fig.~\ref{Fig6} yields
A=(23.2$\pm$0.5)\,cm$^{-2}$K$^{-1}$ and $T_{C}$= (-242$\pm$10)\K.
However, the fit is of low quality since the experimental
$\omega_{SM}(T)$ saturates at low temperatures below 50\K.
Therefore it is more justified to use the Barrett
formula\cite{minaki03}
\begin{equation}\label{Minaki}
 \omega_{SM}(T)=\sqrt{A\Big[\Big(\frac{T_{1}}{2}\Big)\coth\Big(\frac{T_{1}}{2T}\Big)-{T_{0}\Big]}},
\end{equation}
which takes into account the quantum fluctuation at low
temperatures. The fit with Eq.~\ref{Minaki} is much more accurate
especially at low temperatures (see Fig.~\ref{Fig6}) and yields
the following parameters: A=(24.0$\pm$0.7)\,cm$^{-2}$K$^{-1}$,
$T_{0}$= (-221$\pm$13)\K\, and $T_{1}$=(113$\pm$14)\K. $T_{1}$ is
the temperature below which the quantum fluctuations start to play
a role ($\frac{1}{2}k_{B}T_{1}$ is the zero-point vibration
energy). In other words, $\omega_{SM}(T)$ follows the Cochran law
above $T_{1}$, while below this temperature $\omega_{SM}(T)$
deviates from this law due to the quantum fluctuations. $T_{0}$ is
the critical temperature, which is in our case negative indicating
the tendency to hypothetical lattice instability at negative
temperatures. Note that the $A$ parameters, as well as the $T_{c}$
and $T_{0}$ parameters in Cochran and Barret fits have the same
values within the experimental errors.

Up to now discussed IR and THz spectra were taken above 6\K, i.e.
above $T_{N}$. Since the permittivity decreases below $T_{N}$, we
may expect an increase in $\omega_{SM}$ in the AFM phase and its
softening with applied magnetic field. In a different setup we
therefore measured the IR reflectivity at 1.8\K\, from 30 to
350\cm\, in the magnetic field from 0 to 13\,T (see
Fig.~\ref{Fig7}). Unfortunately, due to the low IR signal, high
noise and limited accuracy of the measurement, we do not see any
reliable magnetic field dependence of our FIR reflectivity. The
expected change of $\omega_{SM}$ with the highest magnetic field
is only 3\cm, which lies below limit of accuracy of the IR
experiment using the magnetic cryostat.

Finally we conclude that our IR experiment confirmed the
predictions of Refs. \cite{katsufuji01} that the quantum
paraelectric behaviour in EuTiO$_{3}$ is caused by a soft optic
phonon and its frequency corresponds to value calculated from the
first principles.\cite{fennie06} Simultaneously, we bring evidence
that the soft mode is coupled with higher frequency polar modes,
predominantly with TO2 phonon. However, the expected change of
$\omega_{SM}$ with magnetic field was not detected due to the low
accuracy of IR reflectivity measurements in the magnetic cryostat.
The effect is expected to be larger in a single crystal or
strained thin films, which will be subject of our next studies.

\acknowledgments The work was supported by the Grant Agency of the Czech Republic
(Project No. 202/06/0403), AVOZ10100520, Ministry of Education (OC101 and OC102-COST539)
and European grant RITA-CT-2003-505474. The authors are grateful to V. Studni\v{c}ka for
the XRD analysis.

\end{document}